# IS-LABEL: an Independent-Set based Labeling Scheme for Point-to-Point Distance Querying on Large Graphs

Ada Wai-Chee Fu, Huanhuan Wu
Dept. of Computer Science and Engineering
The Chinese University of Hong Kong
adafu,hhwu@cse.cuhk.edu.hk

James Cheng, Shumo Chu
School of Computer Engineering
Nanyang Technological University, Singapore
j.cheng,shumo.chu@acm.org

Raymond Chi-Wing Wong
Dept. of Computer Science and Engineering
Hong Kong University of Science & Technology
raywong@cse.ust.hk

## ABSTRACT

We study the problem of computing shortest path or distance between two query vertices in a graph, which has numerous important applications. Quite a number of indexes have been proposed to answer such distance queries. However, all of these indexes can only process graphs of size barely up to 1 million vertices, which is rather small in view of many of the fast-growing real-world graphs today such as social networks and Web graphs. We propose an efficient index, which is a novel labeling scheme based on the independent set of a graph. We show that our method can handle graphs of size three orders of magnitude larger than those existing indexes.

## 1. INTRODUCTION

Computing the shortest path or distance between two vertices is a basic operation in processing graph data. The importance of the operation is not only because of its role as a key building block in many algorithms but also of its numerous applications itself. In addition to applications in transportation, VLSI design, urban planning, operations research, robotics, etc., the proliferation of network data in recent years has introduced a broad range of new applications. For example, social network analysis, page similarity measurement in Web graphs, entity relationship ranking in semantic Web ontology, routing in telecommunication networks, context-aware search in social networking sites, to name but a few.

In many of these new applications, however, the size of the underlying graph is often in the scale of millions to billions of vertices and edges. Such large graphs are becoming more and more common, some of the well-known ones include Web graphs, various social networks (e.g., Twitter, Facebook, LinkedIn), RDF graphs, mobile phone networks, SMS networks, etc. Computing shortest path or distance in these large graphs with conventional algorithms such as Dijkstra's algorithm or simple BFS may result in a long running time that is not acceptable.

For computing shortest path or distance between two points in a road network, many efficient indexes have been proposed [1, 2, 3, 8, 13, 14, 26, 27, 28]. However, these works apply unique properties of road networks and hence are not applicable for other graphs/networks that are not similar to road networks. In recent years, a number of indexes have been proposed to process distance queries in general sparse graphs [10, 12, 13, 17, 30, 32, 33]. However, as we will discuss in details in Section 3, these indexes can only handle relatively small graphs due to high index construction cost and large index storage space. As a reference, the largest real graphs tested in these works have only 581K vertices with average degree 2.45 [10], and 694K vertices with average degree 0.45 [17], while most of the other real graphs tested are significantly smaller.

We propose a new index for computing shortest path or distance between two query vertices and our method can handle graphs with hundreds of millions of vertices and edges. Our index, named as **IS-LABEL**, is designed based on a novel application of the *independent set* of a graph, which allows us to organize the graph into layers that form a hierarchical structure. The hierarchy can be used to guide the shortest path computation and hence leads to the design of effective vertex labels (i.e., the index) for distance computation.

We highlight the main contributions of our paper as follows.

- We propose an efficient index for answering shortest path or distance queries, which can handle graphs up to three orders of magnitude larger than those tested in the existing works [10, 12, 13, 17, 30, 32, 33]. None of these existing works can handle even the medium-sized graphs that we tested.

- We design an effective labeling scheme such that the label size remains small even if no optimization (mostly NP-hard) is applied as in the existing labeling schemes.

- Our index naturally lends itself to the design of simple and efficient algorithms for both index construction and query processing.

- We develop I/O-efficient algorithms to construct the vertex labels in large graphs that may not fit in main memory.

- We verify both the efficiency and scalability of our method for processing distance queries in large real-world graphs.

**Organization.** Section 2 defines the problem and basic notations. Section 3 discusses the limitations of existing works. Sections 4 and 5 present the details of index design, and Section 6 describes the algorithms. Section 7 reports the experimental results. Section 8 discusses various issues such as handling path queries, directed graphs, and update maintenance. Section 9 concludes the paper.

Table 1: Frequently-used notations

| Notation | Description |
|---|---|
| $G = (V_G, E_G, \omega_G)$ | A weighted, undirected simple graph |
| $|G| = (|V_G| + |E_G|)$ | The size of $G$ |
| $\omega_G(u, v)$ | The weight of an edge $(u, v)$ in $G$ |
| $adj_G(v)$ | The set of adjacent vertices of $v$ in $G$ |
| $SP_G(u, v)$ | A shortest path from $u$ to $v$ in $G$ |
| $dist_G(u, v)$ | The distance from $u$ to $v$ in $G$ |

## 2. NOTATIONS

We focus our discussion on weighted, undirected simple graphs. Let $G = (V_G, E_G, \omega_G)$ be such a graph, where $V_G$ is the set of vertices, $E_G$ is the set of edges, and $\omega_G : E_G \to \mathbb{N}^+$ is a function that assigns to each edge a positive integer as its weight. We denote the weight of an edge $(u, v)$ by $\omega(u, v)$. The size of $G$ is defined as $|G| = (|V_G| + |E_G|)$.

We define the set of *adjacent* vertices (or *neighbors*) of a vertex $v$ in $G$ as $adj_G(v) = \{u : (u, v) \in E_G\}$, and the *degree* of $v$ in $G$ as $deg_G(v) = |adj_G(v)|$.

We assume that a graph is stored in its adjacency list representation (whether in memory or on disk), where each vertex is assigned a unique vertex ID and vertices are ordered in ascending order of their vertex IDs.

Given a path $p$ in $G$, the *length* of $p$ is defined as $len(p) = \sum_{e \in p} \omega_G(e)$, i.e., the sum of the weights of the edges on $p$. Given two vertices $u, v \in V_G$, the *shortest path* from $u$ to $v$, denoted by $SP_G(u, v)$, is a path in $G$ that has the minimum length among all paths from $u$ to $v$ in $G$. We define the *distance* from $u$ to $v$ in $G$ as $dist_G(u, v) = len(SP_G(u, v))$. We define $dist_G(v, v) = 0$ for any $v \in V_G$.

**Problem definition:** we study the following problem: given a graph $G = (V_G, E_G, \omega_G)$, construct a *disk-based index* for processing *point-to-point* (**P2P**) shortest path or distance queries, i.e., given any pair of vertices $(s, t) \in (V_G \times V_G)$, find $dist_G(s, t)$.

We focus on *sparse graphs*, since most large and many fast growing real-world networks are sparse. We will focus our discussion on processing P2P distance queries. Computing the actual path will be a fairly simple extension with some extra bookkeeping, which will be discussed in Section 8, where we will also show that our index can be extended to handle directed graphs.

Table 1 gives the frequently-used notations in the paper.

## 3. LIMITATIONS OF EXISTING WORK

We highlight the challenges of computing P2P distance by discussing existing approaches and their limitations.

### 3.1 Indexing Approaches

Cohen et al. [13] proposed the 2-hop labeling that computes for each vertex $v$ two sets, $L_{in}(v)$ and $L_{out}(v)$, where for each vertex $u \in L_{in}(v)$ and $w \in L_{out}(v)$, there is a path from $u$ to $v$ and from $v$ to $w$. The distances $dist_G(u, v)$ and $dist_G(v, w)$ are pre-computed. Given a distance query, $s$ and $t$, the index ensures that $dist_G(s, t)$ can be answered as $\min_{v \in (L_{out}(s) \cap L_{in}(t))} \{dist_G(s, v) + dist_G(v, t)\}$. However, computing the 2-hop labeling, including the heuristic algorithms [12, 30], is very costly for large graphs. Moreover, the size of the 2-hop labels is too big to be practical for large graphs.

Xiao et al. [33] exploit symmetric structures in an unweighted undirected graph to compress BFS trees to answer distance queries. However, the overall size of all the compressed BFS trees is prohibitively large even for medium sized graphs.

Wei [32] proposed an index based on a tree decomposition of an undirected graph $G$, where each node in the tree stores a set of vertices in $G$. The distance between each pair of vertices stored in each tree node is pre-computed, so that queries can be answered by considering the minimum distance between vertices stored in a simple path in the tree. However, the pair-wise distance computation for vertices stored in the tree nodes, especially in the root node, is expensive and requires huge storage space. As a result, the method cannot scale to handle large graphs.

Recently Chang et al. [10] also applied tree decomposition to compute multi-hop labels that trade query efficiency of 2-hop labels [13] for indexing cost. Similar to [32], tree decomposition is an expensive operation and the graphs that can be handled by their method are still relatively small.

Jin et al. [17] proposed to use a spanning tree as a highway structure in an directed graph, so that distance from $s$ to $t$ is computed as the length of the shortest path from $s$ to some vertex $u$, then from $u$ via the highway (i.e., a path in the spanning tree) to some vertex $v$, and finally from $v$ to $t$. Every vertex is given a label so that a set of entry points in the highway (e.g., $u$) and a set of exit points (e.g., $v$) can be obtained. However, the labeling is too costly, in terms of both time and space, for the method to be practical for even medium sized graphs (e.g., one step in the process requires all pairs shortest paths to be computed and input to another step).

The problem of P2P distance querying has been well studied for road networks. Abraham et al. [2] recently proposed a hub-based labeling algorithm, which is the fastest known algorithm in the road network setting. This method incorporates heuristical steps in distance labeling by making use of the concepts of contraction hierarchies [14] and shortest path covers [13]. There are other fast algorithms such as [27], [14], and [8], that are also based on the concept of a hierarchy of highways to reduce the search space for computing shortest paths. However, it has been shown in [3] and [1] that the effectiveness of these methods relies on properties such as low VC dimensions and low highway dimensions, which are typical in road networks but may not hold for other types of graphs. Another approach is based on a concise representation of all pairs shortest paths [26, 28]. However, this approach heavily depends on the spatial coherence of vertices and their inter-connectivity. Therefore, while P2P distance querying has been quite successfully resolved for road networks, these methods are in general not applicable to graphs from other sources.

Cheng et al. [11] proposed an index for computing the distance from a source vertex to all other vertices, which can be used to compute P2P distance, but much computation will be wasted in computing the distances from the source to many irrelevant vertices.

### 3.2 Other Approaches

When the input graph is too large to fit in main memory, external memory algorithms can be used to reduce the high disk I/O cost. Existing external memory algorithms are mainly for computing single-source shortest paths [18, 22, 23, 20, 21] or BFS [5, 6, 9, 19, 24], which are wasteful for computing P2P distance. In addition, external memory algorithms are very expensive in practice.

There are also a number of approximation methods [7, 15, 25, 29, 31] proposed to compute P2P distance. Although these methods have a lower complexity than the exact methods in general, they are still quite costly for processing large graphs, in terms of both preprocessing time and storage space. We focus on exact distance querying but remark that approximation can be applied on top of our method (e.g., on the graph $G_k$ defined in Section 5).

# 4. QUERYING DISTANCE BY VERTEX HIERARCHY

In this section, we present our main indexing scheme, which consists of the following components:

- A layered structure of vertex hierarchy constructed from the input graph.
- A vertex labeling scheme developed from the vertex hierarchy.
- Query processing using the set of vertex labels.

We discuss each of these three components in Sections 4.1 to 4.3.

## 4.1 Construction of Vertex Hierarchy

The main idea of our index is to assign hierarchy to vertices in an input graph $G$ so that we can use the vertex hierarchy to compute the vertex labels, which are then used for querying distance.

To create hierarchies for vertices in $G$, we construct a layered hierarchical structure from $G$. To formally define the hierarchical structure, we first need to define the following two important properties that are crucial in the design of our index:

- **Vertex independence**: given a graph $H = (V_H, E_H, \omega_H)$, and a set of vertices $I$, we say that $I$ maintains the vertex independence property with respect to $H$ if $I \subseteq V_H$ and $\forall u, v \in I, (u, v) \notin E_H$, i.e., $I$ is an *independent set* of $H$.

- **Distance preservation**: given two graphs $H_1 = (V_{H_1}, E_{H_1}, \omega_{H_1})$ and $H_2 = (V_{H_2}, E_{H_2}, \omega_{H_2})$, we say that $H_2$ maintains the distance preservation property with respect to $H_1$ if $\forall u, v \in V_{H_2}, dist_{H_2}(u, v) = dist_{H_1}(u, v)$.

While distance preservation is essential for processing distance queries, vertex independence is critical for efficient index construction as we will see later when we introduce the index.

We now formally define the layered hierarchical structure, followed by an illustrating example.

DEFINITION 1 (VERTEX HIERARCHY). *Given a graph $G = (V_G, E_G, \omega_G)$, a **vertex hierarchy** structure of $G$ is defined by a pair $(\mathbb{L}, \mathbb{G})$, where $\mathbb{L} = \{L_1, \ldots, L_h\}$ is a set of vertex sets and $\mathbb{G} = \{G_1, \ldots, G_h\}$ is a set of graphs such that:*

- $V_G = L_1 \cup \ldots \cup L_h$, and $L_i \cap L_j = \emptyset$ for $1 \leq i < j \leq h$;

- *For $1 \leq i \leq h$, each $L_i$ maintains the vertex independence property with respect to $G_i$, i.e., $L_i$ is an independent set of $G_i$;*

- $G_1 = G$, *and for $2 \leq i \leq h$, let $G_i = (V_{G_i}, E_{G_i}, \omega_{G_i})$, then $V_{G_i} = (V_G - L_1 - \ldots - L_{i-1})$, whereas $E_{G_i}$ and $\omega_{G_i}$ satisfy the condition that $G_i$ maintains the distance preservation property with respect to $G_{i-1}$.*

Intuitively, $\mathbb{L}$ is a *partition* of the vertex set $V_G$ and represents a vertex hierarchy, where $L_i$ is at a lower hierarchical level than $L_j$ for $i < j$. Meanwhile, each $G_i \in \mathbb{G}$ preserves the distance information in the original graph $G$, as shown by the following lemma.

LEMMA 1. *For all $u, v \in V_{G_i}$, where $1 \leq i \leq h$, $dist_{G_i}(u, v) = dist_G(u, v)$.*

PROOF. Since for any $u, v \in V_{G_i}$, $u, v \in V_{G_j}$ for $1 \leq j \leq i$. Thus, we have $dist_{G_i}(u, v) = dist_{G_{i-1}}(u, v) = \ldots = dist_{G_1}(u, v) = dist_G(u, v)$ since each $G_i$ maintains the distance preservation property with respect to $G_{i-1}$ for $2 \leq i \leq h$. □

We use the following example to illustrate the concept of vertex hierarchy.

EXAMPLE 1. *Figure 1 shows a given graph $G$ and the vertex hierarchy of $G$. We assume that each edge in $G$ has unit weight except for $(e, f)$, which has a weight of 3. It is obvious that the set $\{c, f, i\}$ forms an independent set in $G$, similarly $\{b, d, h\}$ in $G_2$ and $\{e\}$ in $G_3$. It is easy to see that $G_2$ preserves all distances in $G$, we shall explain the addition of edge $(e, h)$ later. In order to preserve the distance in $G_2$, an edge $(e, g)$ of weight 2 is added to $G_3$. $G_4$ consists of a single edge $(a, g)$ of weight 3. $L_4 = \{a\}$, $G_5$ consists of a single vertex $g$, $L_5 = \{g\}$.*

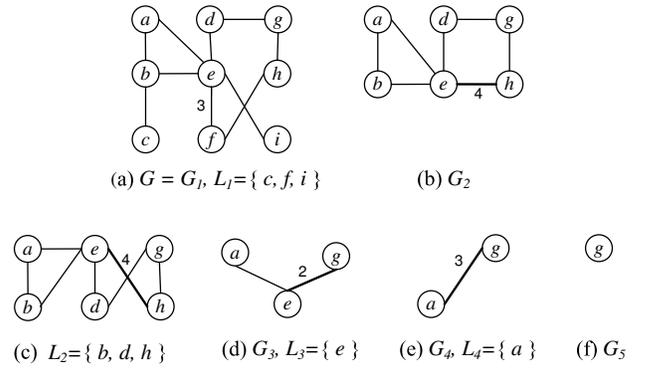

(a) $G = G_1$, $L_1 = \{c, f, i\}$  (b) $G_2$

(c) $L_2 = \{b, d, h\}$  (d) $G_3$, $L_3 = \{e\}$  (e) $G_4$, $L_4 = \{a\}$  (f) $G_5$

**Figure 1: A vertex hierarchy**

The distance preservation property can be maintained in $G_i$ with respect to $G_{i-1}$ as follows. First, we require the subgraph of $G_{i-1}$ induced by the vertex set $V_{G_i}$ to be in $G_i$ (i.e. $(u, v) \in E_{G_i}$ iff $(u, v) \in E_{G_{i-1}}$ for $u, v \in V_{G_i}$). Then, we create a set of additional edges, called **augmenting edges**, to be included into $E_{G_i}$ as follows. For any vertex $v \in L_{i-1}$ (thus $v \notin V_{G_i}$ according to Definition 1), if $u, w \in V_{G_i}$, $(u, v) \in E_{G_{i-1}}$ and $(v, w) \in E_{G_{i-1}}$, then an augmenting edge $(u, w)$ is created in $G_i$ with $\omega_{G_i}(u, w) = \omega_{G_{i-1}}(u, v) + \omega_{G_{i-1}}(v, w)$. If $(u, w)$ already exists in $G_i$, then $\omega_{G_i}(u, w) = \min(\omega_{G_{i-1}}(u, w), \omega_{G_{i-1}}(u, v) + \omega_{G_{i-1}}(v, w))$. An edge in $G_i$ with updated weight is also called an augmenting edge. For example, in Figure 1, in $G_3$, $dist(e, g)$ can be preserved by creating an augmenting edge $(e, g)$ with $\omega(e, g) = 2$. Edge $(e, h)$ is also added according to our process above. Note that $dist_{G_1}(e, h) = 3$, which can be preserved in $G_2$ without adding $(e, h)$, but we leave $(e, h)$ there to avoid costly distance querying needed to exclude $(e, h)$.

The following lemma shows the correctness of constructing $G_i$ from $G_{i-1}$ as discussed above.

LEMMA 2. *Constructing $G_i$ from $G_{i-1}$, where $2 \leq i \leq h$, by adding augmenting edges to the induced subgraph of $G_{i-1}$ by $V_{G_i}$, maintains the distance preservation property with respect to $G_{i-1}$.*

PROOF. According to Definition 1, $L_{i-1}$ is the only set of vertices that are in $G_{i-1}$ but missing in $G_i$. For any two vertices $s$ and $t$ in $G_i$, suppose that the shortest path (in $G_{i-1}$) from $s$ to $t$, $SP_{G_{i-1}}(s, t)$ does not pass through any vertex in

$L_{i-1}$, then the distance between $s$ and $t$ in $G_{i-1}$ is trivially preserved in $G_i$. Next suppose $SP_{G_{i-1}}(s,t)$ passes through some vertex $v \in L_{i-1}$. Let $SP_{G_{i-1}}(s,t) = \langle s, \ldots, u, v, w, \ldots, t \rangle$. Then, we must have the augmenting edge $(u, w)$ created in $G_i$ with $\omega_{G_i}(u,w) = \omega_{G_{i-1}}(u,v) + \omega_{G_{i-1}}(v,w)$, or $\omega_{G_i}(u,w) = \min(\omega_{G_{i-1}}(u,w), \omega_{G_{i-1}}(u,v) + \omega_{G_{i-1}}(v,w))$ if $(u,w)$ already exists in $G_i$. Therefore, the distance (in $G_{i-1}$) between any two vertices is preserved in $G_i$. □

In addition to the distance preservation property that is required for answering distance queries, the proof also gives a hint on why we require each $L_i$ to be an independent set of $G_i$. Since there is no edge in $G_{i-1}$ between any two vertices in $L_{i-1}$, to create an augmenting edge $(u,w)$ in $G_i$ we only need to do a self-join on the neighbors of the vertex $v \in L_{i-1}$. Thus, the search space is limited to 2 hops from each vertex. On the contrary, if an edge can exist between two vertices in $L_{i-1}$, then to preserve the distance the search space is at least 3 hops from each vertex, which is significantly larger than the 2-hop search space in practice. This is crucial for processing a large graph that cannot fit in main memory as we may need to scan the graph many times to perform the join, as we will see in Section 6.

## 4.2 Vertex Labeling

With the vertex hierarchy $(\mathbb{L}, \mathbb{G})$, we now describe a labeling scheme that can facilitate fast computation of P2P distance. We first define the following concepts necessary for the labeling.

- **Level number**: each vertex $v \in V_G$ is assigned a level number, denoted by $\ell(v)$, which is defined as $\ell(v) = i$ iff $v \in L_i$.

- **Ancestor**: a vertex $u \in V_G$ is an ancestor of a vertex $v$ if there exists a sequence $S = \langle v = w_1, w_2, \ldots, w_p = u \rangle$, such that $\ell(w_1) < \ell(w_2) < \ldots < \ell(w_p)$, and for $1 \leq i < p$, the edge $(w_i, w_{i+1}) \in E_{G_j}$ where $j = \ell(w_i)$. Note that $v$ is an ancestor of itself. If $u$ is an ancestor of $v$, then $v$ is a **descendant** of $u$.

EXAMPLE 2. *In our example in Figure 1, the level numbers of $c, f, i$ are 1, that of $b, d, h$ are 2, that of $e$ is 3. The ancestors of $f$ will be $e, h, a, g$, since $(f,e)$ and $(f,h)$ are in $G_1$, $(h,g)$ is in $G_2$, and $(e,a), (e,g)$ are in $G_3$. Note that $d$ is not an ancestor of $f$ since in the path $\langle f, e, d \rangle$, $\ell(e) = 3$ while $\ell(d) = 2$. The ancestor-descendant relationships are shown in Figure 2(a).*

We now define vertex label as follows.

DEFINITION 2 (VERTEX LABEL). *The **label** of a vertex $v \in V_G$, denoted by $LABEL(v)$, is defined as $LABEL(v) = \{(u, dist_G(v,u)) : u \in V_G \text{ is an ancestor of } v\}$.*

To compute $LABEL(v)$ for all $v \in V_G$, we need to compute the distance from $v$ to each of $v$'s ancestors. This is an expensive process which cannot be scaled to process large graphs. To address this problem, we define a relaxed vertex label that requires only an upper-bound, $d(v,u)$, of $dist_G(v,u)$ and show that $d(v,u)$ suffices for answering distance queries.

DEFINITION 3 (RELAXED VERTEX LABEL). *The **relaxed label** of a vertex $v \in V_G$, denoted by $label(v)$, is a set of "$(u, d(v,u))$" pairs computed by the following procedure: For each $v \in V_G$, we first include $(v, 0)$ in $label(v)$ and mark $v$. Then, we add more entries to $label(v)$ recursively as follows. Take a marked vertex $u$ that has the smallest level number $\ell(u)$, and unmark $u$. Let $\ell(u) = j$. For each $w \in adj_{G_j}(u)$, where $\ell(w) > j$ and $(w, d(v,w)) \notin label(v)$, add the entry $(w, (d(v,u) + \omega_{G_j}(u,w)))$ to $label(v)$, and mark $w$. If the entry $(w, d(v,w))$ is already in $label(v)$, update $d(v,w) = \min(d(v,w), (d(v,u) + \omega_{G_j}(u,w)))$. Repeat the above recursive process until no more vertex is marked.*

As for $LABEL(v)$, $label(v)$ contains entries for all ancestors of $v$. In Section 6, we will show that the new definition facilitates the design of an I/O-efficient algorithm for handling large graphs. Here, we further illustrate the concept using an example, and then prove that $label(v)$ can indeed be used instead of $LABEL(v)$ to correctly answer P2P distance queries in the following subsection.

EXAMPLE 3. *For our example in Figure 1, the ancestor relationships are shown in Figure 2(a), where all edges have unit weights unless indicated otherwise. The labeling starts with $L_1$, for vertices $c, f, i$, next $L_2$ vertices $b, d, h$ are labeled, followed by $L_3 = \{e\}$, $L_4 = \{a\}$, and $L_5 = \{g\}$. Consider the labeling for vertex $c$, first, $(c, 0)$ is included, since $adj_G(c) = \{b\}$, $(b, 1)$ is added to $label(c)$ and $b$ is marked. $b$ is unmarked by checking its neighbors $a$ and $e$ in $G_2$, and we include both $(a, 2), (e, 2)$ into $label(c)$, $a$ and $e$ are marked. $e$ is at level 3 and is unmarked next. $adj_{G_3}(e) = \{a, g\}$, we add $(g, 4)$ to $label(c)$. Then $a$ is unmarked, its only neighbor $g$ in $G_4$ is already in $label(c)$, $d(c,g)$ is not updated. $g$ is marked. Finally $g$ is unmarked, since $g$ has no neighbor in $G_5$, no further processing is required. The labels for all vertices are shown in Figure 2(b). Note that $d(h,e) = 4$ in $label(h)$, while $dist_G(h,e) = 3$, hence $d(h,e) > dist_G(h,e)$. In general the distance value in a label entry can be greater than the true distance.*

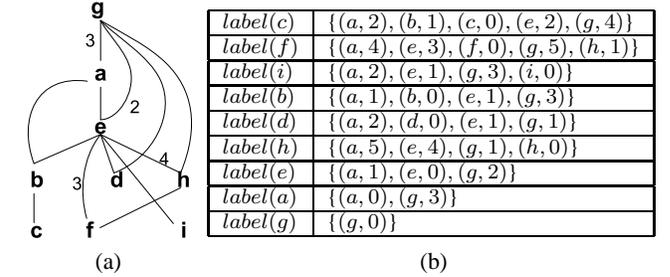

**Figure 2: Labeling for the example in Figure 1**

## 4.3 P2P Distance Querying

We now discuss how we use the vertex labels to answer P2P distance queries. We first define the following label operations used in query processing.

- **Vertex extraction**: $\mathcal{V}[label(v)] = \{u : (u, d(v,u)) \in label(v)\}$.

- **Label intersection**: $label(u) \cap label(v) = \mathcal{V}[label(u)] \cap \mathcal{V}[label(v)]$.

The above two operations apply in the same way to $LABEL(.)$. Given a P2P distance query with two input vertices, $s$ and $t$, let $\mathbb{X} = label(s) \cap label(t)$, the query answer is given as follows.

$$dist_G(s,t) = \begin{cases} \min_{w \in \mathbb{X}}\{d(s,w) + d(w,t)\} & \text{if } \mathbb{X} \neq \emptyset \\ \infty & \text{if } \mathbb{X} = \emptyset \end{cases} \quad (1)$$

In Equation 1, we retrieve $d(s,w)$ and $d(t,w)$ for each $w \in \mathbb{X}$ from $label(s)$ and $label(t)$, respectively. We give an example of answering P2P distance queries using the vertices labels as follows.

EXAMPLE 4. *Consider the example in Figure 1, the labeling is shown in Figure 2. Suppose we are interested in $dist_G(h, e)$. We look up $label(h)$ and $label(e)$. $label(h) \cap label(e) = \{e, a, g\}$. Among these vertices, $g$ has the smallest sum of $d(h, g) + d(g, e) = 1 + 2 = 3$. Hence we return 3 as $dist_G(h, e)$. Note that although the distance $d(h, e)$ recorded in $label(h)$ is 4, which is greater than $dist_G(h, e)$, the correct distance is returned. If we want to find $dist_G(a, g)$, $label(a) \cap label(g) = \{g\}$. Hence $dist_G(a, g)$ is given by $d(a, g) + d(g, g) = 3 + 0 = 3$.*

Query processing using the vertex labels is simple; however, it is not straightforward to see how the answer obtained is correct for every query. In the remainder of this section, we prove the correctness of the query answer obtained using the vertex labels.

We first define the concept of **max-level vertex**, denoted by $v_{max}$, of a shortest path, which is useful in our proofs. Given a shortest path from $s$ to $t$ in $G$, $SP_G(s, t) = \langle s = v_1, v_2, \ldots, v_p = t \rangle$, $v_{max}$ is the max-level vertex of $SP_G(s, t)$ if $v_{max}$ is a vertex on $SP_G(s, t)$ and $\ell(v_{max}) \geq \ell(v_i)$ for $1 \leq i \leq p$. The following lemma shows that $v_{max}$ is unique in any shortest path.

LEMMA 3. *Given two vertices $s$ and $t$, if $SP_G(s, t)$ exists, then there exists a unique max-level vertex, $v_{max}$, of $SP_G(s, t)$.*

PROOF. First, since $SP_G(s, t)$ exists, $v_{max}$ must exist on $SP_G(s, t)$. Now suppose to the contrary that $v_{max}$ is not unique, i.e., there exists at least one other vertex $v$ on $SP_G(s, t)$ such that $\ell(v_{max}) = \ell(v) = j$, which also means that both $v_{max}$ and $v$ are in $L_j$ and $G_j$. Since $L_j$ is an independent set of $G_j$, there is no edge between $v_{max}$ and $v$ in $G_j$. Since $v_{max}$ and $v$ are on the same path $SP_G(s, t)$, they must be connected in $G_j$ and the path connecting them must pass through some neighbor $u$ of $v_{max}$ or $v$ in $G_j$, where $u$ is also on $SP_G(s, t)$. Thus, $u$ cannot be in $L_j$ (otherwise the vertex independence property is violated) and hence $\ell(u) > \ell(v_{max})$, which contradicts that $v_{max}$ is the max-level vertex of $SP_G(s, t)$. □

Next we prove that $LABEL(.)$ can be used to correctly answer P2P distance queries. Then, we show how $label(.)$ possesses the essential information of $LABEL(.)$ for the processing of distance queries.

THEOREM 1. *Given a P2P distance query with two input vertices, $s$ and $t$, let $\mathbb{X} = LABEL(s) \cap LABEL(t)$, then $dist_G(s, t) = \min_{w \in \mathbb{X}} \{dist_G(s, w) + dist_G(t, w)\}$ if $\mathbb{X} \neq \emptyset$, or $dist_G(s, t) = \infty$ if $\mathbb{X} = \emptyset$.*

PROOF. We first show that if $SP_G(s, t)$ exists, then $v_{max} \in \mathbb{X}$. Consider a sequence of vertices, $S = \langle s = u_1, u_2, \ldots, u_\alpha = v_{max} = v_\beta, \ldots, v_2, v_1 = t \rangle$, extracted from $SP_G(s, t)$, such that $\ell(u_1) < \ell(u_2) < \ldots < \ell(u_\alpha) = \ell(v_{max}), \ell(v_1) < \ell(v_2) < \ldots < \ell(v_\beta) = \ell(v_{max})$, and for $1 \leq i < \alpha$, any vertex $w$ between $u_i$ and $u_{i+1}$ on $SP_G(s, t)$ has $\ell(w) < \ell(u_i)$, and same for any vertex between $v_i$ and $v_{i+1}$. Note that since $u_{i+1}$ is the next vertex after $u_i$ with $\ell(u_{i+1}) > \ell(u_i)$, we have $\ell(w) \leq \ell(u_i)$, and $\ell(w) \neq \ell(u_i)$ by the vertex independence property.

Since $u_i$ and $u_{i+1}$ are connected, they must exist together in $G_{\ell(u_i)}$. Since there exists no other vertex $w$ between $u_i$ and $u_{i+1}$ on $SP_G(s, t)$ such that $\ell(w) \geq \ell(u_i)$, $u_i$ and $u_{i+1}$ are not connected by any such $w$ in $G_{\ell(u_i)}$. Thus, by Lemma 1, the edge $(u_i, u_{i+1})$ must exist in $G_{\ell(u_i)}$ for $G_{\ell(u_i)}$ to preserve the distance between $u_i$ and $u_{i+1}$, which means that for $1 \leq j \leq \alpha$, $u_j$ is an ancestor of $s$ and hence $u_j \in LABEL(s)$. Note that $u_1 = s \in LABEL(s)$ if $\alpha = 1$. Similarly, we have $v_i \in LABEL(t)$, for $1 \leq i \leq \beta$. Thus, $v_{max} = u_\alpha = v_\beta \in \mathbb{X}$ and hence $dist_G(s, t) = dist_G(s, v_{max}) + dist_G(t, v_{max})$.

The other case is that $SP_G(s, t)$ does not exist, i.e., $s$ and $t$ are not connected, and we want to show that $\mathbb{X} = \emptyset$. Suppose on the contrary that there exists $w \in \mathbb{X}$. Then, it means that there is a path from $s$ to $w$ and from $t$ to $w$, implying that $s$ and $t$ are connected, which is a contradiction. Thus, $\mathbb{X} = \emptyset$ and $dist_G(s, t) = \infty$ is correctly computed. □

Theorem 1 reveals two pieces of information that are essential for answering distance queries: the ancestor set and the distance to the ancestors maintained in $LABEL(.)$. We first show that $label(.)$ also encodes the same ancestor set of $LABEL(.)$.

LEMMA 4. *For each $v \in V_G$, $\mathcal{V}[label(v)] = \mathcal{V}[LABEL(v)]$.*

PROOF. First, we show that if $w \in \mathcal{V}[LABEL(v)]$, i.e., $w$ is an ancestor of $v$, then $w \in \mathcal{V}[label(v)]$. According to the definition of ancestor, there exists a sequence $S = \langle v = w_1, w_2, \ldots, w_p = w \rangle$, such that $\ell(w_1) < \ell(w_2) < \ldots < \ell(w_p)$, and for $1 \leq i < p$, $(w_i, w_{i+1}) \in E_{G_{\ell(w_i)}}$. This definition implies that if $w_i$ is currently in $\mathcal{V}[label(v)]$, $w_{i+1}$ will also be added to $\mathcal{V}[label(v)]$ according to Definition 3. Since $w_1 = v$ must be in $\mathcal{V}[label(v)]$, it follows that $w = w_p$ is also in $\mathcal{V}[label(v)]$.

Next, we show that if $w \in \mathcal{V}[label(v)]$, then $w \in \mathcal{V}[LABEL(v)]$. First, we have $v \in \mathcal{V}[label(v)]$, $v$ is also in $\mathcal{V}[LABEL(v)]$. Then, according to Definition 3, a vertex $w$ is added to $\mathcal{V}[label(v)]$ only if $w \in adj_{G_{\ell(u)}}(u)$ for some $u$ currently in $\mathcal{V}[label(v)]$, and $\ell(w) > \ell(u)$, and since $u$ is an ancestor of $v$, it implies that $w$ is an ancestor of $v$ and hence $w \in \mathcal{V}[LABEL(v)]$. □

Next, we show that $label(.)$ also possesses the essential distance information for correct computation of P2P distance.

LEMMA 5. *Given a P2P distance query, $s$ and $t$, let $\mathbb{X} = label(s) \cap label(t)$. If $SP_G(s, t)$ exists, then $v_{max} \in \mathbb{X}$, $d(s, v_{max}) = dist_G(s, v_{max})$ and $d(t, v_{max}) = dist_G(t, v_{max})$.*

PROOF. It follows from Lemma 4 that $label(s) \cap label(t) = LABEL(s) \cap LABEL(t)$. As the proof of Theorem 1 shows that $v_{max} \in LABEL(s) \cap LABEL(t)$, we also have $v_{max} \in \mathbb{X}$.

The proof of Theorem 1 defines a sequence, $S = \langle s = u_1, u_2, \ldots, u_\alpha = v_{max} = v_\beta, \ldots, v_2, v_1 = t \rangle$, extracted from $SP_G(s, t)$. In particular, the proof shows that the edge $(u_i, u_{i+1})$ exists in $G_{\ell(u_i)}$ and $\ell(u_{i+1}) > \ell(u_i)$, for $1 \leq i < \alpha$. Thus, according to Definition 3, we add the entry $(u_{i+1}, (d(s, u_i) + \omega_{G_{\ell(u_i)}}(u_i, u_{i+1})))$ to $label(s)$. Since each $\omega_{G_{\ell(u_i)}}(u_i, u_{i+1})$ preserves the distance between $u_i$ and $u_{i+1}$, and $d(s, u_1) = dist_G(s, u_1)$, it follows that $d(s, v_{max} = u_\alpha) = dist_G(s, v_{max} = u_\alpha)$. Similarly, we have $d(t, v_{max}) = dist_G(t, v_{max})$. □

Finally, the following theorem states the correctness of query processing using $label(.)$.

THEOREM 2. *Given a P2P distance query, $s$ and $t$, $dist_G(s, t)$ evaluated by Equation 1 is correct.*

PROOF. The proof follows directly from Theorem 1, Lemmas 4 and 5. □

## 5. A K-LEVEL VERTEX HIERARCHY

In Definition 1, we do not limit the height $h$ of the vertex hierarchy, i.e., the number of levels in the hierarchy. This definition ensures that an independent set $L_i$ can always be obtained for each $G_i$, for $1 \leq i \leq h$. However, there are two problems associated with the height of the vertex hierarchy. First, as the number of levels $h$ increases, the label size of the vertices at the lower levels (i.e., vertices with a smaller level number) also increases. Since vertex labels require storage space and are directly related to query processing, there is a need to limit the vertex label size. Second, as we will discuss in Section 6, the complexity of constructing the vertex hierarchy is linear in $h$. Thus, reducing $h$ can also improve the efficiency of index construction.

In this section, we propose to limit the height $h$ by a $k$-level vertex hierarchy, where $k$ is normally much smaller than $h$, and discuss how the above-mentioned problems are resolved.

## 5.1 Limiting the Height of Vertex Hierarchy

The main idea is to terminate the construction of the vertex hierarchy earlier at a level when certain condition is met. We first define the $k$-level vertex hierarchy.

DEFINITION 4 (K-LEVEL VERTEX HIERARCHY). *Given a graph $G = (V_G, E_G, \omega_G)$, a vertex hierarchy structure $\mathbb{H} = (\mathbb{L}, \mathbb{G})$ of $G$, and an integer $k$, where $1 < k \leq (h+1)$ and $h$ is the number of levels in $\mathbb{H}$, a **k-level vertex hierarchy** structure of $G$ is defined by a pair $(\mathbb{H}_{<k}, G_k)$, where $\mathbb{H}_{<k}$ and $G_k$ are defined as follows:*

- $\mathbb{H}_{<k} = (\mathbb{L}_{<k}, \mathbb{G}_{<k})$ *consists of the first $(k-1)$ levels of $\mathbb{H}$, i.e., $\mathbb{L}_{<k} = \{L_1, \ldots, L_{k-1}\}$ and $\mathbb{G}_{<k} = \{G_1, \ldots, G_{k-1}\}$;*
- $G_k$ *is the same $G_k$ as the $G_k$ in $\mathbb{G}$.*

The $k$-level vertex hierarchy simply takes the first $(k-1)$ $L_i \in \mathbb{L}$, for $1 \leq i < k$, and the first $k$ $G_i \in \mathbb{G}$, for $1 \leq i \leq k$. We set the value of $k$ as follows: let $i$ be the first level such that $(|G_i|/|G_{i-1}|) > \sigma$, where $\sigma$ ($0 < \sigma \leq 1$) is a threshold for the effect of $G_i$; then, $k = i$.

If $k = (h+1)$, then $\mathbb{H}_{<k}$ is simply $\mathbb{H}$ and $G_k$ is an empty graph. In practice, a value of $\sigma$ that attains a reasonable indexing cost and storage usage will often give $k \ll h$.

For the $k$-level vertex hierarchy, we assign the level number $\ell(v) = i$ for each vertex $v \in L(i)$, where $1 \leq i \leq (k-1)$, while for each vertex $v \in V_{G_k}$, we assign $\ell(v) = k$. In this way, we can compute $label(v)$ (or $LABEL(v)$) for each vertex $v \in V_G$ in the same way as discussed in Section 4.2. Note that $label(v) = \{(v, 0)\}$ for each vertex $v \in V_{G_k}$ since $v$ has the highest level number among all vertices in $V_G$.

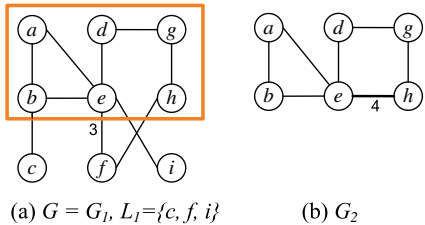

**Figure 3: A $k$-level vertex hierarchy ($k = 2$)**

EXAMPLE 5. *Let us consider our running example in Figure 1, if we set $k = 2$, there is only one level $L_1$ in $\mathbb{L}_{<k}$, the graph $G_2$ is the highest level graph and is not further decomposed. The $k$-level vertex hierarchy is shown in Figure 3. The maximum level of vertices is 2, since all vertices $v$ in $G_2$ are assigned $\ell(v) = 2$. The labels for the vertices in $L_1$ are shown in the following table.*

| $label(c)$ | $\{(b,1),(c,0)\}$ |
|---|---|
| $label(f)$ | $\{(e,3),(f,0),(h,1)\}$ |
| $label(i)$ | $\{(e,1),(i,0)\}$ |

## 5.2 P2P Distance Querying by k-Level Vertex Hierarchy

According to Section 5.1, $\ell(v)$ and $label(v)$ computed from the $k$-level vertex hierarchy may be different from those computed from the original vertex hierarchy. However, we show later in this section that these labels are highly useful for they capture all the information that is essential from $G - G_k$ for a continued distance search in $G_k$. Given a P2P distance query, $s$ and $t$, we process the query according to whether $s$ and $t$ are in $G_k$. We have the following two possible types of queries.

**Type 1**: $s \notin V_{G_k}$ and $t \notin V_{G_k}$, and either $(\mathcal{V}[label(s)] \cap V_{G_k}) = \emptyset$ or $(\mathcal{V}[label(t)] \cap V_{G_k}) = \emptyset$. Type 1 queries are evaluated by Equation 1.

**Type 2**: queries that are not Type 1. Type 2 queries are evaluated by a *label-based bi-Dijkstra search* procedure.

We have discussed query processing by Equation 1 in Section 4.3. We now discuss how we process Type 2 queries as follows.

### 5.2.1 Label-based bi-Dijkstra Search

We describe a bidirectional Dijkstra's algorithm that utilizes vertex labels for effective pruning. The algorithm consists of two main stages: (1) initialization of distance queues and pruning condition, and (2) bidirectional Dijkstra search.

As shown in Algorithm 1, we first initialize a *forward* and a *reverse* min-priority queue, *FQ* and *RQ*, which are to be used for running Dijkstra's single-source shortest path algorithm from $s$ and $t$, respectively. For any vertex $v \in V_{G_k}$, if $(v, d(s,v)) \in label(s)$, we add $(v, d(s,v))$ to *FQ* with $d(s,v)$ as the key. For all other vertices in $V_{G_k}$ but not in $label(s)$, we add the record $(v, \infty)$ to *FQ*. Similarly, we initialize *RQ*.

The vertex labels can also be used for pruning the search space. If there exists a path between $s$ and $t$ that passes through some vertex $w \in (V_G - V_{G_k} - \{s, t\})$, then Lines 5-6 initializes $\mu$ as the minimum length of such a path. Note that $\mu \geq dist_G(s, t)$.

We now describe Stage 2 of the query processing. We run Dijkstra's algorithm simultaneously from $s$ and $t$ by extracting the vertex $v$ with the minimum key from *FQ* or *RQ* (Line 9). Let $(v, d(x, v))$ be the extracted record, where $x = s$ if the record is extracted from *FQ* and $x = t$ otherwise. At this point, Dijkstra's algorithm guarantees that the distance from $x$ to $v$ is found, i.e., $d(x, v) = dist_G(x, v)$. Then, in Lines 13-18, the distance from $x$ to every neighbor $u$ of $v$ in $G_k$ is updated, if $u$ is still in *FQ* (if $x = s$) or *RQ* (if $x = t$).

In addition to starting the search in both directions from $s$ and $t$ in Dijkstra's algorithm, we also add a pruning condition in Line 8 that requires the sum of the minimum keys of *FQ* and *RQ* to be less than $\mu$. If this sum is not less than $\mu$, then it means that no path from $s$ to $t$ of a shorter distance than $\mu$ can be found (proved in Theorem 4) and hence we return $dist_G(s, t) = \mu$.

To improve the pruning effect so as to converge the search quickly, we keep updating $\mu$ whenever $d(x, u)$ is updated if $dist_G(x', u)$ has been found (Lines 17-18), since $u$ is a potential vertex on $SP_G(s, t)$. We use a set $S$ to keep a set of vertices whose distance from $s$ or $t$ has been found. Whenever $dist_G(x, v)$ is found for a vertex $v$, if $v$ is not yet in $S$, we insert $v$, together with $dist_G(x, v)$, into $S$.

We give an example to illustrate how queries are processed as follows.

EXAMPLE 6. *Let us consider Example 5. Suppose we need to process a distance query between vertices $c$ and $i$, i.e. $s = c$, $t = i$. In $label(c)$, $b$ is in $G_k$, and therefore we enter $(b, d(c, b) = 1)$*

**Algorithm 1:** Label-based bi-Dijkstra Search

**Input** : $s, t, label(s), label(t), G_k$
**Output** : $dist_G(s,t)$
// Stage 1: initialization of distance queues and pruning condition
// FQ (RQ): forward (reverse) min-priority queue

1. initialize $FQ$ with the set $\{(v, d(s,v)) : v \in V_{G_k}, (v, d(s,v)) \in label(s)\}$, with $d(s,v)$ as the key;
2. initialize $RQ$ with the set $\{(v, d(t,v)) : v \in V_{G_k}, (v, d(t,v)) \in label(t)\}$, with $d(t,v)$ as the key;
3. $\forall v \in V_G$ and $v$ not in $FQ(RQ)$, insert $(v, \infty)$ into $FQ(RQ)$;
// $\mu$: shortest distance from $s$ to $t$ found so far
// $\mu$ is used for pruning in Stage 2
4. $\mu \leftarrow \infty$;
5. $\mathbb{X} \leftarrow label(s) \cap label(t)$;
6. **if** $\mathbb{X} \neq \emptyset$ **then** $\mu \leftarrow \min_{w \in \mathbb{X}}\{d(s,w) + d(w,t)\}$;

// Stage 2: bidirectional Dijkstra search
7. $S \leftarrow \emptyset$;
8. **while** *both FQ and RQ are not empty, and* $(\min(FQ) + \min(RQ)) < \mu$ **do**
9.     $(v, d(x,v)) \leftarrow$ *extract-min(FQ, RQ)*; // $x = s$ or $x = t$
10.     let $x' = t$ if $x = s$, and $x' = s$ if $x = t$;
11.     **if** $\langle v, dist_G(x,v)\rangle$ *is not in* $S$ **then**
12.       insert $\langle v, dist_G(x,v)\rangle$ into $S$;
13.     **foreach** $u \in adj_{G_k}(v)$ **do**
14.       **if** $d(x,u) > d(x,v) + \omega_{G_k}(v,u)$ **then**
15.         $d(x,u) \leftarrow d(x,v) + \omega_{G_k}(v,u)$;
16.         update $d(x,u)$ in $FQ$ (if $x = s$) or $RQ$ (if $x = t$);
17.         **if** $\langle u, dist_G(x',u)\rangle$ *is in* $S$ **then**
18.           $\mu \leftarrow \min\{\mu, d(x,u) + dist_G(x',u)\}$;

19. **return** $\mu$;

into $FQ$. In $label(i)$, $e$ is in $G_k$, hence we enter $(e, d(i,e) = 1)$ into $RQ$. $label(c) \cap label(i) = \phi$, hence $\mu = \infty$ after Stage 1 of Algorithm 1. In Stage 2, let us extract $(b, 1)$ from $FQ$ first, $\langle b, 1\rangle$ is inserted into $S$, and we enter $(a, 2)$, $(e, 2)$, into $FQ$. Next we extract $(e, 1)$ from $RQ$, and insert $\langle e, 1\rangle$ into $S$. $(a, 2)$, $(d, 2)$, $(b, 2)$ are entered into $RQ$. Since $b$ is in $S$, we update $\mu$ to $2 + 1 = 3$. At this point $(min(FQ) + min(RQ)) > \mu$ and we return $dist_G(c,i) = 3$.

### 5.2.2 Correctness

We now prove the correctness of query processing by the $k$-level vertex hierarchy. We first prove the correctness for processing Type 1 queries.

THEOREM 3. *Given a P2P distance query, s and t, if the query belongs to Type 1, then $dist_G(s,t)$ evaluated by Equation 1 is correct.*

PROOF. First, we show that if the query belongs to Type 1, then $SP_G(s,t)$ does not contain any vertex in $V_{G_k}$. Suppose on the contrary that $SP_G(s,t)$ contains a vertex in $V_{G_k}$. Then, consider the sub-path of $SP_G(s,t)$ from $s$ to $x$, where $x$ is the only vertex on the sub-path that is in $V_{G_k}$. Since $SP_G(s,t)$ is a shortest path in $G$, this sub-path is a shortest path from $s$ to $x$ in $G$. Let $SP_G(s,x)$ be the sub-path. Consider the query with two input vertices $s$ and $x$; then, by similar argument as in the proof of Lemma 3 we have $v_{max} = x$ on $SP_G(s,x)$, and by similar argument as in the proof of Lemma 5 we have $x = v_{max} \in \mathcal{V}[label(s)]$. A symmetric analysis on the sub-path from $t$ to some vertex $y$, where $y$ is the only vertex on the sub-path that is in $V_{G_k}$, shows that $y = v_{max}$ on $SP_G(t,y)$ and $y \in \mathcal{V}[label(t)]$. This contradicts the definition of Type 1 query that either $(\mathcal{V}[label(s)] \cap V_{G_k}) = \emptyset$ or $(\mathcal{V}[label(t)] \cap V_{G_k}) = \emptyset$.

Now if $SP_G(s,t)$ does not contain any vertex in $V_{G_k}$, then the query can be answered using only label entries of vertices from the first $(k-1)$ levels of the vertex hierarchy. These entries will have identical occurrences and contents in the vertex labels at the first $k$ levels of any vertex hierarchy $\mathbb{H}_{<j}$, where $k \leq j \leq h+1$, which is formed by limiting the height of a given $\mathbb{H}$. Thus, the correctness of query answer follows from Theorem 2. □

Note that Type 1 queries exist only if there exist more than one connected component in $G$ such that all vertices in some connected component(s) have a level number lower than $k$.

Next we prove the correctness for processing Type 2 queries.

THEOREM 4. *Given a P2P distance query, s and t, if the query belongs to Type 2, then $dist_G(s,t)$ evaluated by the label-based bi-Dijkstra search procedure is correct.*

PROOF. We have two cases: (1) $SP_G(s,t)$ does not contain any vertex in $V_{G_k}$, or (2) otherwise.

If $SP_G(s,t)$ does not contain any vertex in $V_{G_k}$, then $dist_G(s,t)$ is computed in Lines 5-6 of Algorithm 1, or in other words by Equation 1. As explained in the proof of Theorem 3, the correctness of query answer follows from Theorem 2.

If $SP_G(s,t)$ contains at least one vertex in $V_{G_k}$, then consider the two subpaths, $SP_G(s,x)$ and $SP_G(t,y)$, defined in the proof of Theorem 3 (note that it is possible $s = x$ and/or $x = y$ and/or $y = t$). $dist_G(s,x)$ and $dist_G(t,y)$ can be answered using only label entries of vertices in $\mathbb{L}_{<k}$ and their ancestors in $G_k$ for $(\mathbb{H}_{<k}, G_k)$. From the labeling mechanism, the occurrences and contents of such label entries will be identical in the labels of vertices in the first $k$ levels of any vertex hierarchy $\mathbb{H}_{<j}$, $k \leq j \leq h+1$, which is formed by limiting the height of a given $\mathbb{H}$. Hence by Theorem 2, $dist_G(s,x)$ and $dist_G(t,y)$ are correctly initialized in Lines 1-3 of Algorithm 1. Thus, if we do not consider the pruning condition in Line 8, then Dijkstra's algorithm guarantees the distance from $s$ (and $t$) to any vertex in $G_k$ correctly computed, from which we can obtain $dist_G(s,t)$.

Now we consider query processing with pruning. Let $\mu = \mu*$, and $min_f = \min(FQ)$ and $min_r = \min(RQ)$, when the search stops. If $\mu*$ is the value of $\mu$ initialized in Line 6, then we must have $x = y \in (label(s) \cap label(t))$ and hence $\mu* = (dist_G(s,x) + dist_G(t,x))$. Otherwise, $\mu*$ is a value assigned to $\mu$ in Line 18 and suppose to the contrary that there exists a shorter path between $s$ and $t$ with length $p$ such that $p < \mu*$. Since the path passes through vertices in $G_k$, there must exist an edge $(v,u)$ in $G_k$ such that $p = dist_G(s,v) + \omega_{G_k}(v,u) + dist_G(u,t)$, $dist_G(s,v) < min_f$ and $dist_G(u,t) < min_r$. The existence of this edge is guaranteed because $p < \mu* \leq (min_f + min_r)$. Since $dist_G(s,v) < min_f$ and $dist_G(u,t) < min_r$, by Dijkstra's algorithm, both $dist_G(s,v)$ and $dist_G(t,u)$ have been computed when the search stops. Thus, $\mu$ should have been updated to a value not greater than $p$ in Line 18 when the edge $(v,u)$ was processed. This contradicts our assumption and hence $\mu* = dist_G(s,t)$. □

## 6. ALGORITHMS

In this section, we present the algorithms for index construction (i.e., vertex hierarchy construction and vertex labeling) and query processing using the vertex labels. In recent years, due to the proliferation of many massive real world networks, there has been an increasing interest in algorithms that handle large graphs. For processing large graphs that cannot fit in main memory, I/O cost usually dominates. Thus, we propose I/O-efficient algorithms, from which the in-memory algorithms can also be easily devised.

For the analysis of the I/O complexity in this section, we define the following notation [4]. Let $scan(N) = \Theta(N/B)$ and $sort(N) = \Theta(\frac{N}{B} \log_{M/B} \frac{N}{B})$, where $N$ is the amount of data being read or written from/to disk, $M$ is the main memory size, and $B$ is the disk block size ($1 \ll B \leq M/2$).

## 6.1 Algorithm for Index Construction

Although the vertex hierarchy, except $G_k$, is not required for query processing, it is needed for vertex labeling. There are two components, $\mathbb{L}$ and $\mathbb{G}$, in the vertex hierarchy; thus, we have the following two main steps: (1) computing each independent vertex set $L_i \in \mathbb{L}$, and (2) constructing each distance-preserving graph $G_i \in \mathbb{G}$. We first describe these two steps, followed by the construction of the overall vertex hierarchy, and finally the vertex labeling.

### 6.1.1 Constructing $L_i$

We want to maximize the size of each $L_i$ as this helps to minimize the number of levels $h$ and hence also minimizes the vertex label size. However, maximizing $L_i$ means computing the maximum independent set of $G_i$, which is an NP-hard problem.

We adopt a greedy strategy to approximate the set of maximum independent set of $G_i$ by selecting the vertex with minimum degree at each step [16], since small degree vertices have smaller number of dependent (i.e., adjacent) vertices and hence more vertices are left as candidates for independent set at the next step. Moreover, the greedy algorithm can also be easily extended to give an I/O-efficient algorithm that handles the case when $G_i$ is too large to fit in main memory, as described in Algorithm 2.

The algorithm computes an independent set $L_i$ of $G_i$, together with the adjacency lists of the vertices in $L_i$, denoted by $ADJ(L_i)$. We use $ADJ(L_i)$ to construct $G_{i+1}$ in Section 6.1.2. To compute $L_i$, we also keep those vertices that have been excluded from $L_i$ in the algorithm, as denoted by $L'$. We use a buffer to keep the current $L_i$ and $ADJ(L_i)$, and another buffer to keep $L'$.

The algorithm first makes a copy of $G_i$, let it be $G'_i$, and then sorts the adjacency lists in $G'_i$ in ascending order of the vertex degrees (i.e., the sizes of the adjacency lists). Then, we read $G'_i$ in this sorted order, i.e., the adjacency lists of vertices with smaller degrees are read first. For each $adj_{G'_i}(u)$ read, if $u$ is not in $L'$, we include $u$ into $L_i$ and add $adj_{G'_i}(u)$ to $ADJ(L_i)$. Meanwhile, we exclude all vertices in $adj_{G'_i}(u)$ from $L_i$ because of their dependence with $u$, i.e., we add these vertices to $L'$. The algorithm terminates when $adj_{G'_i}(u)$ for all $u$ in $G'_i$ are read.

If $G_i$ is very large, it is possible that $L_i$ and $ADJ(L_i)$ are too large to be kept by a memory buffer. We can simply write the current $L_i$ and $ADJ(L_i)$ in the buffer to disk, and then clear the buffer for new contents of $L_i$ and $ADJ(L_i)$. However, when the buffer for $L'$ is full, we cannot simply flush the buffer since it is possible that $\exists u \in L'$, $adj_{G'_i}(u)$ has not been read yet. To tackle this without incurring random disk accesses, we scan $G'_i$ to remove all the vertices currently in $L'$, together with their adjacency lists, from $G'_i$, because these vertices have already been excluded from $L_i$. Then, we clear the buffer for $L'$.

If $G'_i$ can be resident in main memory, Lines 10-11 of Algorithm 2 are not necessary and we only need to scan $G'_i$ once. If $G'_i$ is resident on disk, it is easy to see that only sequential scans of $G'_i$ are needed and expensive random disk access is avoided.

Algorithm 2 takes $sort(|G_i|)$ I/Os to sort $G_i$. If $|L'| < M$, we need another $scan(|G_i|)$ I/Os to read $G_i$. Otherwise, $O(|L'|/M) * scan(|G_i|)$ I/Os are required.

---

**Algorithm 2:** Constructing $L_i$

**Input** : A graph $G_i = (V_{G_i}, E_{G_i}, \omega_{G_i})$
**Output** : $L_i$ and $ADJ(L_i) = \{adj_{G_i}(v) : v \in L_i\}$

1 allocate a buffer for $L_i$ and $ADJ(L_i)$, and a buffer for $L'$;
2 $G'_i \leftarrow G_i$;
3 sort $adj_{G'_i}(v)$ in $G'_i$ in ascending order of $deg_{G'_i}(v)$;
4 **foreach** $adj_{G'_i}(u)$ read in $G'_i$ **do**
5     **if** $u \notin L'$ **then**
6         insert $u$ into $L_i$, and insert $adj_{G'_i}(u)$ into $ADJ(L_i)$;
7         **foreach** $v \in adj_{G'_i}(u)$ **do**
8             **if** $v \notin L'$ **then** insert $v$ into $L'$;
9     **if** *buffer for $L_i$ and $ADJ(L_i)$ is full* **then** flush the buffer;
10    **if** *buffer for $L'$ is full* **then**
11        scan $G'_i$ to delete all $v \in L'$ and $adj_{G'_i}(v)$, and clear $L'$;

---

**Algorithm 3:** Constructing $G_i$

**Input** : $G_{i-1}$, $L_{i-1}$ and $ADJ(L_{i-1})$
**Output** : $G_i$

1 $G_i \leftarrow G_{i-1}$;
2 remove from $G_i$ all $v \in L_{i-1}$ and $adj_{G_{i-1}}(v)$;
3 $E_A \leftarrow \emptyset$;
4 **foreach** $adj_{G_{i-1}}(v) \in ADJ(L_{i-1})$ **do**
5     **foreach** $u, w \in adj_{G_{i-1}}(v)$, *where $u < w$* **do**
6         insert into $E_A$ the edges $(u, w)$ and $(w, u)$, with $\omega_{G_i}(u, w) = \omega_{G_i}(w, u) = (\omega_{G_{i-1}}(u, v) + \omega_{G_{i-1}}(v, w))$;
7 sort the edges in $E_A$ by vertex ID's;
8 scan $E_A$ and $G_i$ to add each edge $(u, w) \in E_A$ to $G_i$, or update $\omega_{G_i}(u, w)$ with the smaller weight if $(u, w)$ already exists in $G_i$;

---

### 6.1.2 Constructing $G_i$

After obtaining $L_{i-1}$ and $ADJ(L_{i-1})$, we use them to construct $G_i$. As shown in Algorithm 3, we first initialize $G_i$ by removing the occurrences of all vertices in $L_{i-1}$, together with their adjacency lists, from $G_{i-1}$. However, the resultant $G_i$ may not satisfy the distance preservation property. As discussed in Section 4.1, the violation to this property can be fixed by the creation of a set of augmenting edges. We create these augmenting edges from $ADJ(L_{i-1})$ as follows.

When a vertex $v \in L_{i-1}$, together with $adj_{G_{i-1}}(v)$, is removed from $G_{i-1}$ to form $G_i$, what is missing in $G_i$ is the path $\langle u, v, w \rangle$ for any $u, w \in adj_{G_{i-1}}(v)$, where $u < w$ (i.e., $u$ is ordered before $w$). Thus, to preserve the distance we only need to create the augmenting edge $(u, w)$, and symmetrically $(w, u)$ for undirected graphs, with weight $(\omega_{G_{i-1}}(u, v) + \omega_{G_{i-1}}(v, w))$.

We create all such augmenting edges in Lines 4-6 of Algorithm 3 and store them in an array $E_A$. Then, we sort the edges in $E_A$ first in ascending order of the first vertex and then of the second vertex. Then, we scan both $E_A$ and $G_i$ (already sorted in its adjacency list representation), so that each edge in $E_A$ is merged into $G_i$. If an edge in $E_A$ is already in $G_i$, then its weight updated to the smaller value of its weight recorded in $E_A$ and in $G_i$.

If main memory is not sufficient, Line 2 of Algorithm 3 uses $O(|L_{i-1}|/M) * scan(|G_{i-1}|)$ I/Os, Lines 3-6 and 8 use $scan(|G_i|)$ I/Os, and Line 7 uses $sort(|G_i|)$ I/Os, since $|E_A| < |G_i|$.

### 6.1.3 Constructing $(\mathbb{L}, \mathbb{G})$

The overall scheme to construct the vertex hierarchy, $(\mathbb{L}, \mathbb{G})$, is to start with the given $G_1 = G$, and keep repeating the two steps

**Algorithm 4:** Top-Down Vertex Labeling

**Input** : $(\mathbb{L}, \mathbb{G})$
**Output** : $label(v), \forall v \in V_G$
```
// Initialization of vertex labels
```
1 **for** $i = 1, ..., k - 1$ **do**
2     **foreach** $v \in L_i$ **do**
3         $label(v) \leftarrow \{(v, 0)\} \cup \{(u, \omega_{G_i}(v, u)) : u \in adj_{G_i}(v)\}$;

4 $\forall v \in V_{G_k}:$  $label(v) \leftarrow \{(v, 0)\}$;

```
// Top-down vertex labeling
```
5 **for** $i = k - 1, ..., 1$ **do**
6     allocate buffer $B_L$ and load $label(v)$, for each $v \in L_i$, in $B_L$;
7     allocate buffer $B_U$ and load $label(v)$, for each $v \in L_j$ for $i < j < k$ and for each $v \in V_{G_k}$, in $B_U$;
8     **foreach** block $B_L$ **do**
9         **foreach** block $B_U$ **do**
10             **foreach** $label(v)$ in $B_L$ **do**
11                 **foreach** $label(u)$ in $B_U$ **do**
12                     **if** $(u, d(v, u)) \in label(v)$ **then**
13                         **foreach** $(w, d(u, w)) \in label(u)$ **do**
14                             **if** $(w, d(v, w)) \notin label(v)$ **then**
15                                 add $(w, d(v, u) + d(u, w))$ to $label(v)$;
16                         **else**
17                                 $d(v, w) = \min(d(v, w), d(v, u) + d(u, w))$;

of computing $L_i$ (Algorithm 2) and constructing $G_i$ (Algorithm 3) until we reach a level $k$ (see Section 5.1 for the value of $k$).

### 6.1.4 Top-Down Vertex Labeling

Definition 3 essentially defines a procedure for computing $label(v)$ for each $v \in V_G$. However, a careful analysis will show that such a procedure, if implemented directly as it is described, involves much redundant processing as implied by the following corollary of Lemma 4.

COROLLARY 1. *Given a vertex* $v \in L_i$, *we have* $\mathcal{V}[label(v)] = \{v\} \cup (\bigcup_{u \in adj_{G_i}(v)} \mathcal{V}[label(u)])$.

PROOF. By Definition 3, $\forall u \in adj_{G_i}(v)$, $u$ will be included into $\mathcal{V}[label(v)]$. From the result of Lemma 4, we have $\forall u \in \mathcal{V}[label(v)]$, $u$ is an ancestor of $v$ by Definition 2. In the same way, we have $\forall w \in \mathcal{V}[label(u)]$, $w \in \mathcal{V}[label(v)]$ since $w$ is then also an ancestor of $v$. Thus, $\forall u \in adj_{G_i}(v)$, $\mathcal{V}[label(u)] \subseteq \mathcal{V}[label(v)]$.

Next, $\forall w \in \mathcal{V}[label(v)]\setminus\{v\}$, $w \in \mathcal{V}[label(u)]$ for some $u \in adj_{G_i}(v)$ because $w$ is included into $\mathcal{V}[label(v)]$ from some $u$ by Definition 3, and by the same procedure $w$ will be included into $\mathcal{V}[label(u)]$ when we compute $label(u)$. □

Corollary 1 implies that $label(v)$ can be computed from $label(u)$, for each $u \in adj_{G_i}(v)$, instead of from scratch. Based on this, we design a more efficient top-down algorithm for vertex labeling as shown in Algorithm 4.

The algorithm consists of two stages: initialization of vertex labels and top-down vertex labeling by block nested loop join, discussed as follows.

According to Corollary 1, we only need to add $(v, 0)$ and $(u, \omega_{G_i}(v, u))$ for all $u \in adj_{G_i}(v)$ to $label(v)$, and then derive other entries of $label(v)$ from $label(u)$ in the top-down process.

For each $v \in V_{G_k}$, however, we only need to add $(v, 0)$ to $label(v)$ since each $v \in V_{G_k}$ has only one ancestor, i.e., $v$ itself.

After the initialization, we compute the labels for the vertices starting from the top levels to the bottom levels, i.e., from level $(k-1)$ down to level 1. We assume that the set of labels at each level may not be able to fit in main memory and hence use block nested loop join to find the matching labels, i.e., $label(u)$ for each $u \in adj_{G_i}(v)$ when we process $v$ at level $i$. Note that if $u \in adj_{G_i}(v)$, then $(u, d(v, u)) \in label(v)$ by the initialization. Thus, as shown in Lines 11-16, we derive the entries of other ancestors of $v$ from $label(u)$ directly, which essentially follows the rule specified in Definition 3.

The complexity of the algorithm is apparently dominated by the top-down process. Let $b_L(i) = |\{label(v) : v \in L_i\}|$, and $b_U(i) = |\bigcup_{i<j<k}\{label(v) : v \in L_j\} \cup \{label(v) : v \in V_{G_k}\}|$. The I/O complexity for the block nested loop join is given by $(b_L(i)/M) * (b_U(i)/B)$. Thus, the I/O complexity of Algorithm 4 is given by $O(\sum_{i=1}^{k-1}((b_L(i)/M) * (b_U(i)/B)))$.

## 6.2 Algorithm for Query Processing

For processing large datasets, the vertex labels may not fit in main memory and are stored on disk. The entries in each $label(v)$ are stored sequentially on disk and are sorted by the vertex ID's of the ancestors of $v$. Thus, $label(s) \cap label(t)$ involves simple sequential scanning of the entries in $label(s)$ and $label(t)$. From our experiments, the vertex labels are small in size and retrieving a vertex label from disk takes only one I/O. The CPU time for query processing comes mostly from the bi-Dijkstra search. For a graph $G = (V, E)$, a binary heap can be used and Dijkstra's algorithms runs in $O((|E| + |V|) \log |V|)$ time.

## 7. EXPERIMENTAL EVALUATION

We evaluate the performance of our method and compare with other related methods for processing P2P distance queries. All systems tested were programmed in C++ and compiled with the same compiler. All experiments were performed on a computer with an Intel 3.3 GHz CPU, using 4GB RAM and a 7200 RPM SATA hard disk, running Ubuntu 11.04 Linux OS.

We use the following datasets in our experiments: Web, BTC, as-Skitter, wiki-Talk and web-Google. BTC is an unweighted graph, which is a semantic graph converted from the Billion Triple Challenge 2009 RDF dataset (http://vmlion25.deri.ie/), where each vertex represents an object such as a person, a document, and an event, and each edge represents the relationship between two nodes such as "has-author", "links-to", and "has-title". Web (http://barcelona.research.yahoo.net/webspam) is a subgraph of the UK Web graph, where vertices are pages and edges are hyperlinks. The original graph $\vec{G}$ is directed and converted into undirected graph G in this way: if two vertices are reachable from each other within $w$ hops in $\vec{G}$, where $w \in \{1, 2\}$, they have an undirected edge with weight $w$ in G. For there are many connected components in G, we extract the largest connected component for our experiments. As-Skitter is an Internet topology graph from traceroutes run daily in 2005 (http://www.caida.org/tools/measurement/skitter). The wiki-Talk network contains all the users and discussions from Wikipedia till January 2008. Nodes in the network represent users of Wikipedia (http://www.wikipedia.org/) and an undirected edge between node i and node j means that user i has at least edited one talk page of user j or vice versa. In web-Google, nodes represent web pages and hyperlinks between them are represented by undirected edges. It was released for Google Programming Contest in

2002 (http://www.google.com/programming-contest/). We list the datasets in Table 2.

|  | $|V|$ | $|E|$ | Avg. Deg | Max Deg | Disk size |
|---|---|---|---|---|---|
| BTC | 164.7M | 361.1M | 2.19 | 105,618 | 5.6 GB |
| Web | 6.9M | 113.0M | 16.40 | 31,734 | 1.1 GB |
| as-Skitter | 1.7M | 22.2M | 13.08 | 35,455 | 200 MB |
| wiki-Talk | 2.4M | 9.3M | 3.89 | 100,029 | 100 MB |
| Google | 0.9M | 8.6M | 9.87 | 6,332 | 80 MB |

**Table 2: Real datasets**

## 7.1 Results of Index Construction

We first report the results for our index construction. We list the number of levels ($k$), the number of vertices ($|V_{G_k}|$) and edges ($|E_{G_k}|$) of the graph $G_k$, the total label size, and indexing time in Table 3. We set the $k$-selection criterion as follows: when the graph size of $G_{i+1}$ is larger than 95% of the graph size of $G_i$, i.e. when $|V_i| + |E_i| >= 0.95 * (|V_{i+1}| + |E_{i+1}|)$, set $k = i$. This is to say that the independent set $L_i$ has introduced less than 5% of graph size reduction. We shall use 95% as our default threshold.

|  | $k$ | $|V_{G_k}|$ | $|E_{G_k}|$ | Label size | Indexing time (seconds) |
|---|---|---|---|---|---|
| BTC | 6 | 134K | 16.4M | 10.6 GB | 2513.73 |
| Web | 19 | 242K | 14.5M | 13.1 GB | 2274.36 |
| as-Skitter | 6 | 86K | 8.5M | 678.3 MB | 483.65 |
| wiki-Talk | 5 | 14K | 2.4M | 152.5 MB | 239.48 |
| Google | 7 | 87K | 2.5M | 199.5 MB | 35.13 |

**Table 3: Index construction results with threshold 0.95**

It is intuitively that with more levels in the vertex hierarchy, we can get a smaller size for graph $G_k$, bigger label size, and longer indexing time. This in turn affects the query time and we shall have more discussion in the next subsection.

## 7.2 Results of Query Performance

To assess query performance, we randomly generate 1000 queries in each dataset and compute the average query time. The results for our datasets are shown in Table 4. The total time for each query is made up of two parts, the first part Time (a) being the time for retrieving labels for $s$ and $t$ if needed, the second part Time (b) is for the bi-Dijkstra search. We note that Time (a) for the dataset Web is much greater since the label size for Web is much bigger. Although BTC is a very large dataset, the query time is very short and this is due to the low average degree in the graph, which makes the bi-Dijkstra search highly efficient. Note that even though wiki-Talk and Google are much smaller in size, Time (a) is still above 10ms, which is due to the speed of our hard disk, with a benchmark of 10ms per disk I/O. For these datasets, the label sizes are very small, and in fact they can be kept in main memory, in which case we will save the factor of Time (a) in the total time. We call this approach **in-memory IS-LABEL**, or **IM-ISL** for short.

|  | $k$ | Total query time(ms) | Time (a) (ms) | Time (b) (ms) |
|---|---|---|---|---|
| BTC | 6 | 11.55 | 11.47 | 0.08 |
| Web | 19 | 28.02 | 20.08 | 7.94 |
| as-Skitter | 6 | 20.05 | 12.68 | 7.37 |
| wiki-Talk | 5 | 12.22 | 10.85 | 1.37 |
| Google | 7 | 12.97 | 10.37 | 2.60 |

**Table 4: Query time with threshold 0.95: Time (a) denotes the time used for getting the label, Time (b) denotes the time used for bi-Dijkstra search**

Table 5 shows results of different query types using IS-LABEL. There are three types of queries: Type 1: Both $s$ and $t$ are in $G_k$; Type 2: One of $s, t$ id in $G_k$; Type 3: Both $s$ and $t$ are not in $G_k$. We can see that Type 1 query has the shortest average query time for there is no need to lookup the labels, Type 2 query requires the lookup of the label of only one query vertex, and for Type 3 we need to retrieve the labels of both query vertices. The time for running the bi-Dijkstra search on $G_k$ does not vary much for the three types of queries.

|  | $k$ | Query type | Total query time(ms) | Time (a) (ms) | Time (b) (ms) |
|---|---|---|---|---|---|
| BTC | 6 | 1 | 0.08 | 0.0 | 0.08 |
|  |  | 2 | 5.85 | 5.73 | 0.12 |
|  |  | 3 | 9.03 | 8.94 | 0.09 |
| Web | 19 | 1 | 10.40 | 0.0 | 10.40 |
|  |  | 2 | 19.61 | 10.14 | 9.47 |
|  |  | 3 | 29.81 | 20.37 | 9.44 |

**Table 5: Query time for 3 types of queries: time (a) denotes the time used for getting the label, time (b) denotes the time used for bi-Dijkstra search**

When index construction is based on different $k$ values, it will affect the querying time. We list the querying results for graph BTC and Web with different $k$ values in Table 6. The greater $k$ is, the smaller the size of graph $G_k$, which leads to shorter time for the bi-directional dijkstra algorithm. However, the time for scanning labels will increase with the increase of the label size with a larger $k$. Considering all factors, we can conclude that the $k$ values that we have chosen automatically as shown in Table 3 are highly effective.

|  | $k$ | $|V_{G_k}|$ | $|E_{G_k}|$ | Label size | Indexing time(s) | Query time(ms) |
|---|---|---|---|---|---|---|
| BTC | 5 | 167K | 17.2M | 7.2 GB | 1555.24 | 10.45 |
| BTC | 6 | 134K | 16.4M | 10.6 GB | 2513.73 | 11.55 |
| BTC | 7 | 114K | 15.8M | 17.1 GB | 7227.40 | 12.37 |
| Web | 18 | 260K | 15.2M | 12.2 GB | 2115.31 | 30.72 |
| Web | 19 | 242K | 14.5M | 13.1 GB | 2274.36 | 28.02 |
| Web | 20 | 226K | 13.8M | 13.9 GB | 2485.24 | 33.65 |

**Table 6: Index construction time, label size, $G_k$ size and query time with different $k$ values**

|  | $k$ | $|V_{G_k}|$ | $|E_{G_k}|$ | Label size | Indexing time(s) | Query time(ms) |
|---|---|---|---|---|---|---|
| BTC | 5 | 167K | 17.2M | 7.2 GB | 1818.21 | 10.64 |
| Web | 7 | 808K | 31.1M | 1.6 GB | 752.69 | 40.85 |
| as-Skitter | 4 | 160K | 9.3M | 221.9 MB | 246.69 | 18.98 |
| wiki-Talk | 4 | 17K | 2.4M | 99.3 MB | 182.32 | 11.38 |
| Google | 6 | 107K | 2.7M | 127.3 MB | 25.57 | 12.96 |

**Table 7: Index Construction time, label size, $G_k$ size, and query time with threshold 0.9**

To investigate how the $k$-selection criterion may impact the overall performance, we examine another setting where we set $k = i$ when $(|G_i|/|G_{i-1}|) > 90\%$. We list the indexing construction results of using 90% as our threshold in Table 7. We can see that a larger threshold gives rise to smaller $k$ values, which lead to larger sizes for $G_k$, smaller label sizes and shorter indexing times. However, the query time in the case of dataset Web becomes greater, which is a trade-off for the smaller indexing costs. Depending on the available resources and application requirements, the threshold can be tuned to a desirable value. However, it can be noted that we maintain very good query time as we vary the choices of the threshold. This shows that our high quality query performance is a robust behavior.

## 7.3 Comparison with Other Methods

There exist a number of recent works on point-to-point distance querying. The most recent work by Jin et al [17] shows that their method out-performs other state-of-the-art approaches. However, the space requirement of their program exceeds our RAM capacity for the larger datasets, while for our smaller datasets, the indexing time was prohibitively long. Note that their results recorded over 70 hours of labeling time for a small dataset with only 694K vertices and 312K edges [17]. We next tried to compare with the method TEDI in [32]. However, TEDI ran out of memory for each of our datasets due to a very large root node in the tree decomposition.

|  | IS-LABEL | IM-ISL | VC-Index(P2P) | IM-DIJ |
| --- | --- | --- | --- | --- |
| BTC | 11.55 ms | – | 4246.09 ms | – |
| Web | 28.02 ms | – | 31655.77 ms | 430.67 ms |
| as-Skitter | 20.05 ms | 7.15 ms | 3712.33 ms | 23.16 ms |
| wiki-Talk | 12.22 ms | 1.23 ms | 553.94 ms | 9.97 ms |
| Google | 12.97 ms | 2.44 ms | 1285.25 ms | 9.09 ms |

**Table 8: Query time of IS-LABEL, in memory IS-LABEL(IM-ISL), VC-Index (converted for P2P) and IM-DIJ**

|  | Index construction time (seconds) | Index size |
| --- | --- | --- |
| BTC | 6221.44 | 3.1 GB |
| Web | 3544.38 | 3.0 GB |
| as-Skitter | 1013.07 | 486.5 MB |
| wiki-Talk | 52.79 | 137.1 MB |
| Google | 70.37 | 211.3 MB |

**Table 9: Indexing costs for VC-Index**

We find that no known point-to-point distance querying mechanism can handle our data sizes, hence we try to compare with the best related method that can be converted to work for point-to-point querying. The most efficient such method is the VC-Index proposed by Cheng et al in [11]. Since VC-Index is for single source shortest paths queries, we modified the source code to make it work specifically for point to point distance queries by making the program stop once the distance from $s$ to $t$ is found. We compare our method with this converted VC-Index method by taking the average query time over 1000 randomly generated queries. For the datasets that can fit into main memory, we also compare our method with the in-memory bidirectional Dijkstra search (IM-DIJ). We list the average query times in Table 8. In Table 9, we list the indexing costs of VC-Index. From the experimental result, first we notice that in-memory bi-Dijkstra cannot work for the dataset BTC since it exceeds the memory capacity. For the smaller datasets, in-memory IS-LABEL (IM-ISL) is faster than the in-memory bi-Dijkstra method (IM-DIJ), and IS-LABEL is much faster than IM-DIJ for the larger dataset Web. Although VC-Index can handle all the datasets including the case where the data does not fit in main memory, we find that IS-LABEL is many times faster than VC-Index in the query time. The speedup is especially significant for the massive graphs. IS-LABEL is 368 times faster for BTC, and 1130 times faster for Web. Meanwhile, the index construction time of IS-LABEL is also less than that of VC-Index.

## 8. PATH QUERIES, DIRECTED GRAPHS, AND UPDATE MAINTENANCE

In this section, we discuss the extension of our method to answer shortest-path queries and to handle directed graphs. We also briefly discuss how update maintenance can be processed when the input graph is updated dynamically.

### 8.1 Shortest-Path Queries

To answer a P2P shortest-path query, we need to keep some extra information in the vertex labels. When an augmenting edge $(u, w)$ is created in $G_i$ with $\omega_{G_i}(u, w) = \omega_{G_{i-1}}(u, v) + \omega_{G_{i-1}}(v, w)$, we also keep the intermediate vertex $v$ along with the augmenting edge to indicate that the edge represents the path $\langle u, v, w \rangle$. Note that $(u, v)$ and $(v, w)$ are edges in $G_{i-1}$, which in turn can be augmenting edges. In the labeling process, instead of adding the entry $(w, d(u, w))$ to $label(u)$, we also attach the intermediate vertex $v$ (if any) for $(u, w)$. Thus, the entry becomes a triple $(w, d(u, w), v)$ (or $(w, d(u, w), \phi)$, if there is no intermediate vertex). Note that we keep the graph $G_k$, and thus the intermediate vertex of any augmenting edge in $G_k$ is directly attached to the edge.

Given a query, $s$ and $t$, if the query is of Type 1, the answer is determined by two label entries, $(w, d(s, w), v)$ and $(w, d(t, w), v')$. If $v \neq \phi$ (similarly for $v'$), we form two new queries $(s, v)$ and $(v, w)$. In this way, we recursively form queries until the intermediate vertex in a label entry is $\phi$. It is then straightforward to obtain the resulting path by linking all the intermediate vertices. If the query is of Type 2, then the answer is determined by two label entries and a path in $G_k$. The subpaths from the two label entries are derived in the same way as we do for a Type 1 query. The path in $G_k$ is expanded into the original path in $G$ by forming new queries, "$u$ and $v$" and "$v$ and $w$", for any augmenting edge $(u, w)$ with the intermediate vertex $v$. For each such query, the corresponding subpath is obtained as discussed above. The I/O complexity of the overall process is given by $O(|SP_G(s, t)|)$, where $|SP_G(s, t)|$ is the number of edges on $SP_G(s, t)$.

### 8.2 Handling Directed Graphs

To handle directed graphs, we need to modify the vertex hierarchy construction as well as the vertex labeling. Let us use $(u, v)$ to indicate an edge from $u$ to $v$ in this subsection. The concept of independent set can be applied in the same way by simply ignoring the direction of the edges. However, for distance preservation, we create an augmenting edge $(u, w)$ at $G_i$ only if $\exists v \in L_{i-1}$ such that $(u, v), (v, w) \in E_{G_{i-1}}$. We distinguish two types of ancestors for a vertex $v$: **in-ancestors** and **out-ancestors**. The definition of in-ancestors is similar to that of ancestors in undirected graphs, except that we only consider edges from higher-level vertices to lower-level vertices. Analogously, the definition of out-ancestors concerns edges going from lower-level vertices to higher-level vertices.

The labeling needs to handle two directions. For each vertex $v$, we need two types of labels defined as follows. The **in-label** of a vertex $v \in V_G$, denoted by $LABEL_{in}(v)$, is defined as $LABEL_{in}(v) = \{(u, dist_G(u, v)) : u \in V_G \text{ is an in-ancestor of } v\}$. The **out-label** of a vertex $v \in V_G$, denoted by $LABEL_{out}(v)$, is defined as $LABEL_{out}(v) = \{(u, dist_G(v, u)) : u \in V_G \text{ is an out-ancestor of } v\}$.

Given a P2P distance query with two input vertices, $s$ and $t$, we compute $\mathbb{X} = LABEL_{out}(s) \cap LABEL_{in}(t)$ and then answer the query in the same way as given in Equation 1.

### 8.3 Update Maintenance

When the input graph is updated, we want to update the vertex labels incrementally rather than to re-compute them from scratch. We consider the cases where vertices, along with their adjacency lists, are inserted or deleted in the graph. For insertion of a new vertex $u$, we add $u$ to $G_k$. Next we consider each vertex $v$ in the adjacency list $adj_G(u)$ of $u$. If $v$ is in $G_k$, then we simply add the edge $(u, v)$ to $E_{G_k}$ with weight $\omega_G(u, v)$. Otherwise, let $v \in L_i$. We add $(u, \omega_G(u, v))$ to $label(v)$. We also need to add $u$ to the descendants of $v$ (a vertex $w$ is a descendant of $v$ if $v$ is an ancestor of $w$). The descendants of $v$ can be viewed as vertices in a tree rooted at $v$. We traverse this tree so that the entry $(u, d(u, w))$ is added to or modified in $label(w)$, where $w$ is a descendant of $v$, so that the value of $d(u, w)$ is set to or decreased to the accumulated distance of $\omega(u, v) + d(v, v_1), ...d(v_i, w)$, where $v, v_1, ..., w$ is a path in the tree. The I/O complexity is given by the number of descendants of $u$. Next we consider the deletion of a vertex $u$. If $u$ is in $G_k$ and no label of other vertices contains $u$, then $u$ can simply be deleted from the adjacency lists of all its neighbors in $G_k$. Otherwise, we look for the descendants of $u$ and remove the entry of $u$ in the label of each descendant. In this case, the I/O complexity is determined by the number of descendants of $u$. The above lazy update mechanism would have little impact on the query performance for a moderate amount of updates, and we can rebuild the index periodically.

## 9. CONCLUSION

In this paper, we introduce an effective disk-based indexing method named IS-LABEL for distance and shortest path querying in massive graphs. The directed graph version of our method simultaneously solves the fundamental problem of reachability. Given the low costs of IS-LABEL in index construction and querying for both massive undirected and massive directed graphs, we expect our method to handle large graphs for reachability queries.